\documentclass[aps,prl,twocolumn]{revtex4}
\usepackage{epsf}
\begin{document}


\title{Decoherence and the Loschmidt echo} 
\author{F.M. Cucchietti, D.A.R. Dalvit, J.P. Paz and W.H. Zurek} 
\affiliation{Theoretical Division, MS B213, Los Alamos National 
Laboratory, Los Alamos, NM 87545} 
\date{\today} 
 
\begin{abstract} 
Environment--induced decoherence causes entropy increase. It can
be quantified using, e.g., the purity $\varsigma={\rm Tr}\rho^2$. 
When the Hamiltonian of a quantum system is perturbed, its sensitivity 
to such perturbation can be measured by the Loschmidt echo $\bar M(t)$.
It is given by the average squared overlap between the perturbed and 
unperturbed state. We describe the relation between the temporal 
behavior of $\varsigma(t)$ and $\bar M(t)$. In this way we show that 
the decay of the Loschmidt echo can be analyzed using tools developed 
in the study of decoherence. In particular, for systems with a 
classically chaotic Hamiltonian the decay of $\varsigma$ and $\bar M$ 
has a regime where it is dominated by the classical Lyapunov exponents. 
\end{abstract} 

\maketitle 

Environment-induced decoherence is an essential 
ingredient of the quantum--classical transition \cite{DecoRef}.
Its implications for the quantum versions of classically chaotic systems
are especially intriguing, as they go beyond the restoration of the
quantum--classical correspondence. Two of us
discussed this issue for the first time in \cite{ZP94}, presenting  
a surprising result that has been since amply corroborated 
\cite{Sarkar,Patanayak,Monteoli1,Bianucci}:
For a quantum system with a classically chaotic Hamiltonian the 
rate at which information about the initial state is degraded by the
environment becomes independent of the system--environment 
coupling strength. This rate (e.g., the von Neumann entropy production 
rate computed from the reduced density matrix of the system) is set 
by the classical Lyapunov exponents, provided that the coupling 
strength are within a certain (wide) range.
This result has important implications and can be a way to define 
quantum chaos \cite{ZP95}. 
A related but independent way to do this was considered first by Asher 
Peres \cite{Peres} by appealing to a property of quantum systems with a 
classically chaotic counterpart: 
Although quantum dynamics is insensitive to small differences in 
initial conditions, it seems to be highly sensitive to perturbations in 
the Hamiltonian. The similar 
idea of ``hypersensitivity'' was studied in an information theoretic 
framework by Caves and coworkers \cite{CavesSchack}. More recently, 
Levstein, Pastawski and collaborators \cite{PastawskiExp,PastawskiEnt}
experimentally studied sensitivity to perturbations by measuring 
the Loschmidt echo in a many-body spin system. 
Their work underscored the need for a systematic study of various 
aspects of the true nature of the sensitivity displayed by quantum 
systems when their Hamiltonian is perturbed. 
Further analytical \cite{PastawskiJalab} and numerical \cite{Cucchietti} 
work triggered an intense activity that helped clarify some aspects of 
the temporal dependence of the echo \cite{Jacquod,CucLewen,Prosen}. 

The measure of the echo signal is the overlap
between two states that evolve from the same initial wave function $\Psi_0$
under the influence of two Hamiltonians (the unperturbed one, $H_0$, and the 
perturbed one $H_\Delta=H_0+\Delta$). More precisely, when $U_0$ and 
$U_\Delta$ denote the corresponding evolution operators, the echo 
is defined as 
\begin{equation}
M_\Delta(t)=|\langle \Psi_0| U_\Delta^\dagger(t) U_0(t)|\Psi_0\rangle|^2. 
\label{MJ}
\end{equation}
The quantity $\bar M(t)$, obtained by averaging $M_\Delta$ over an 
ensemble of perturbations, can be studied analytically and displays a 
rich temporal dependence. One interesting regime was analyzed by Jalabert and 
Pastawski \cite{PastawskiJalab} who showed, using a semiclassical approximation, 
that there is a window of values for the perturbation strength for 
which $\bar M(t)$ decays with a rate equal to the classical Lyapunov 
exponent. 
In spite of the simple discussion presented above, the 
physically relevant evolution will typically not be unitary: 
Decoherence caused by the interaction with the environment will 
suppress the echo even in the absence of the perturbation $\Delta$.
We shall, however, adhere to the usual assumption 
\cite{PastawskiExp,PastawskiEnt,PastawskiJalab,Cucchietti,Jacquod,CucLewen,Prosen} that the evolutions are 
unitary, and show that even in that case 
of decoherence--free echo supression it is possible to draw useful 
conclusions from the analogy we discuss below.

In this letter we establish a direct connection between decoherence and 
the decay of the Loschmidt echo. In particular, we relate the 
evolution of $\bar M(t)$ and the linear entropy (or purity) of an open 
quantum system. The existence of a kinship between decoherence and the 
decay of the Loschmidt echo is not unexpected: such possibility was 
noted, for example, in Refs. \cite{PastawskiJalab,PastawskiEnt}, but was 
never formally established. Demonstrating the relation of 
these two important areas is interesting not just from a fundamental point
of view but may also prove useful since it allows one to use results 
obtained in the theory of open quantum systems 
to understand better the behavior of the echo. 

The key step in our demonstration is a simple observation: The average echo
$\bar M(t)$, for an ensemble of perturbations characterized by a 
probability density $P(\Delta)$, is
\begin{equation}
\bar M(t)=\int D\Delta\ P(\Delta)\ |\langle\Psi_0|U_\Delta^\dagger (t) 
U_0(t)|\Psi_0\rangle|^2.\label{Mbar1}
\end{equation}
This equation can be rewritten in a more convenient way by defining the 
density matrix of the average perturbed state:
\begin{equation}
\bar\rho(t)=\int D\Delta\ P(\Delta) \ U_\Delta(t)|\Psi_0\rangle
\langle\Psi_0| U_\Delta^\dagger(t).
\label{rhobar}
\end{equation}
In fact, $\bar M(t)$ is simply the overlap between the average state 
$\bar \rho(t)$ and the unperturbed density matrix $\rho_0(t)$ evolved 
from the initial state,  
$\rho_0(t)=U_0(t)|\Psi_0\rangle\langle\Psi_0|U_0^\dagger(t)$:
\begin{equation}
\bar M(t)={\rm Tr}\left(\bar\rho(t)\rho_0(t)\right).
\label{mbarrhobar}
\end{equation}
Once we recognize this simple fact we can go one step further and 
notice that the state $\bar\rho$ generally evolves in time according 
to a master equation which is of the same kind as the ones arising 
in the study of decoherence. Hence, 
the evolution of the echo $\bar M(t)$ is directly placed 
in the context of open quantum systems and decoherence.

Equation (\ref{mbarrhobar}) can be used to establish an 
inequality between the average echo $\bar M(t)$ and the 
purity $\varsigma(t) = {\rm Tr} \bar \rho^2(t)$ (used to characterize 
decoherence). Using Schwartz inequality (applied to the inner product 
between two hermitian operators) and assuming that the initial state is pure
we find that 
\begin{equation}
\bar M^2(t)\le \varsigma(t).
\label{inequality}
\end{equation}
Related inequalities were noticed and used in a somewhat different context  
in \cite{Prosen,Saraceno}. Eq. (\ref{inequality}) implies that when the 
purity ${\rm Tr}\bar \rho^2$ decays exponentially with a rate $\gamma_D$, 
then the overlap $\bar M(t)$ should also decay exponentially (or faster) 
with a rate that should be no less than 
$\gamma_D/2$. However, as we will see later (and as has been established  
in the literature \cite{Monteoli1,PastawskiJalab,Cucchietti}), there is
an important regime (the so--called Lyapunov regime) where both quantities 
decay with the same rate set by the Lyapunov exponent.  

Let us now analyze some generic features of the evolution of the 
average state $\bar\rho(t)$. It is useful to notice that $\bar\rho$ 
generally obeys a master 
equation with non--unitary terms. These terms arise because averaging of the 
evolution over an ensemble of perturbations yields an effect analogous 
(although not equivalent) to the tracing out of the unobserved degrees 
of freedom \cite{DecoRef}. We will find it convenient to consider a 
simple form of perturbation (even though results are not strongly 
dependent on it). Let us assume that $\Delta(x,t)=V(x)J(t)$, 
where $V(x)$ is a function of the coordinates of our system
and $J(t)$ is an external source. For this case, averaging over the 
perturbation $\Delta$ consists of averaging over functions $J(t)$. 
We will also assume that the probability density $P(J)$ is a 
Gaussian functional whose width defines the temporal 
correlation function for the sources. Thus, we will use 
\begin{equation}
P(J)=N\exp\left(-{1\over 2}\int\int dt dt' J(t)\nu^{-1}(t,t')\ J(t')\right), 
\label{PJ}
\end{equation}
where $N$ is a normalization factor. The correlation function of the 
source is then $\int DJ\ P(J) J(t)J(t')=\nu(t,t')$. Using this, we 
can show that the evolution operator for $\bar\rho(t)$ has a path 
integral representation with an influence functional \cite{FV} given by
\begin{equation}
F[x,x']=\exp(-{1\over 2}\int\int dt dt' V_{-}(t) \nu(t,t')\ V_{-}(t')),
\label{IF}
\end{equation}
where $V_{-}(t)=V(x(t))-V(x'(t))$. 
In some simple but physically relevant cases it is possible to write 
a master equation for $\bar\rho(t)$. In fact, when the noise is white, 
i.e. $\nu(t,t')=2D \delta(t-t')$, we can show that 
\begin{equation} 
\dot{\bar\rho}={1\over i\hbar}\left[H_0,\bar\rho \right] 
 -\ D\left[V(x),\left[V(x),\bar\rho\right]\right].
\label{master} 
\end{equation} 
While the first term on the rhs of this equation generates unitary 
evolution, the second term is responsible for decoherence: It induces a 
tendency towards diagonalization in position basis and, in the Wigner 
representation, it gives rise to a diffusion term. For the simplest case 
of $V(x)=x$ the equation for the Wigner function reads
\begin{equation}
\dot W(x,p)=\left\{H_0,W\right\}_{MB}\ +\ D\ \partial^2_{pp}W(x,p),
\label{wignereq}
\end{equation}
where the bracket in the rhs is the so--called Moyal bracket, responsible
for unitary evolution \cite{DecoRef}.

Equations like (\ref{master}) and (\ref{wignereq}) arise if we consider 
a quantum system interacting with a quantum environment formed by a 
collection of harmonic oscillators \cite{CL}. 
In such a case the absolute value of the influence functional generated
by the environment is identical to (\ref{IF}) provided one chooses 
the spectral density and the initial state of the environment in such a 
way that the noise--kernel it produces is equal to the kernel $\nu(t,t')$
appearing in (\ref{IF}). However, in general the 
influence functional is a complex number whose phase is responsible for 
dissipation (noise and dissipation kernels are connected 
as mandated by the fluctuation--dissipation theorem). There is a physically
relevant limit (usually associated with high temperatures) where 
relaxation effects can be ignored. This is the most interesting limit 
in decoherence studies aimed at understanding the quantum--classical 
correspondence \cite{DecoRef}. Thus, in such a limit, the evolution of 
the average state $\bar\rho$ is identical to that of a quantum system
that interacts with an environment.

A convenient way to visualize the transition from quantum to classical 
is provided by the Wigner function, whose oscillations are the signature 
of quantum interference. They should be suppressed by the decoherence term 
to make the quantum--classical correspondence possible. This is 
indeed what happens: When the Wigner function oscillates with a 
well defined wave--vector $k_p$ along the momentum direction 
($W(x,p,t)~\simeq~A(x,t)~\cos{(k_p p)}$), the decoherence term 
in (\ref{wignereq}) washes out oscillations exponentially fast 
with a rate $\Gamma_D=D k_p^2$. We can see the behavior of a 
typical Wigner function for a chaotic system (a driven double well 
analyzed in \cite{Monteoli1}) with and without decoherence in Figure 1. 
Taking into account our previous discussion, the echo $\bar M(t)$ 
is obtained by computing the overlap between the two Wigner functions 
displayed in the figures. The purity 
$\varsigma$ should be computed by taking the overlap of 
the decohered Wigner function with itself. Below, we will discuss 
the relation between these two quantities. 

\begin{figure}
\centering \leavevmode
\epsfxsize 3.2in
\epsfbox{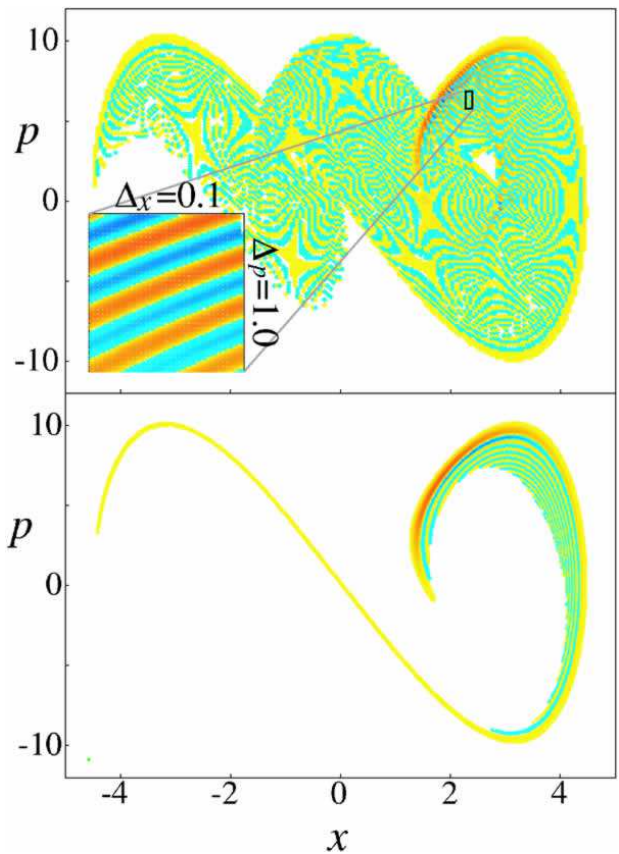}
\caption{Wigner function of an initially Gaussian state evolved with a 
chaotic Hamiltonian without (top) and with (bottom) decoherence. The 
system is a particle moving in a driven double well 
potential (see \cite{Monteoli1}). A region of area $\hbar$ is shown
in the top panel where sub-Planckian structure is evident
\cite{ZurekNature}. The color scale is positive from yellow to red, 
shades of blue are negative and white is zero. In the top panel 
we can appreciate the distinct regions $A_O$ ($A_C$) where 
the Wigner function $W_0$ oscillates rapidly (is positive), used 
in Eq. (\ref{Integral}).}
\label{WignerFigure} 
\end{figure}
 
The master equation (\ref{wignereq}) can be used to obtain 
the time derivatives of the purity $\varsigma$ 
and the echo $\bar M$: 
\begin{eqnarray} 
\dot {\varsigma}&=&
2D\int dxdp\ \bar W(x,p)\partial^2_{pp} \bar W(x,p),
\label{puritydot}\\
\dot{\bar M}&=&
D\int dxdp\ W_0(x,p)\partial^2_{pp}\bar W(x,p).\label{Mdot}
\end{eqnarray} 
Equation (\ref{puritydot}) has been used before to show the existence 
of a domain of exponential decay for the purity ${\rm Tr} \bar \rho^2$ 
\cite{ZP94,ZP95,Patanayak,Monteoli1}
with a rate that, for classically chaotic systems, is independent 
of the diffusion $D$. The central piece of the argument is the 
following: After integrating by parts, equation (\ref{puritydot}) can 
be rewritten as $\dot{\varsigma}/ \varsigma=-2D/\bar\sigma^{2}$, 
where $\bar\sigma$ characterizes the dominant wavelength in the 
spectrum of the Wigner function 
(i.e. $\bar\sigma^{-2}=\int (\partial_p \bar W)^2/ \int \bar W^2$). 
Thus, the rate of change of the purity becomes independent of the 
diffusion constant when $\bar\sigma^{2}$ is proportional to $D$. This 
happens indeed as a consequence of the competition between two effects.  
The first one is the tendency of chaotic evolution to 
generate (exponentially fast, at a rate set by the
Lyapunov exponent $\lambda$) small scale structure in the Wigner function. 
The second effect is due to diffusion, which tends to wash out small scales
exponentially fast at a rate determined by the product $D k_p^2$. As 
demonstrated in \cite{ZP94}, these two effects reach a balance when 
the scale is such that $\bar\sigma^{2}=2D/\lambda$. 
When this equilibrium is reached, the 
purity $\varsigma$ decreases exponentially at a rate fixed by $\lambda$. 
 
For this behavior to take place the diffusion constant should be 
above a threshold \cite{ZP95}. Otherwise the critical width is not 
established (indeed, in the previous argument the implicit assumption 
is that the time scale for diffusion to wash out a $k_p$--oscillation 
is shorter than the time scale for the oscillations to be regenerated
by the dynamics). This simple scenario enables us to 
understand why there is a regime where purity decreases exponentially 
with a Lyapunov rate. The same argument can also be used to analyze 
the exponential decay of the Loschmidt echo. In fact, 
equations (\ref{puritydot}) and (\ref{Mdot}) just differ by a 
factor of $2$ and by the presence 
of $W_0$ instead of $\bar W$ inside the integral. As before, we 
can transform the evolution equation of the echo into 
$\dot{\bar M}/\bar M=-D/\sigma^{2}$, 
$\sigma^{-2}=\int W_0\partial^2_{pp} \bar W/\int W_0 \bar W$. 
When decoherence is effective and the dominant structure in $\bar W$ 
approaches the critical value, the smallest scales of the pure Wigner 
function $W_0$ continue contracting and developing smaller and 
smaller scales  (sub--Planck scales are reached quickly in chaotic
quantum systems \cite{ZurekNature}). In such a case, one can easily 
conclude that $\bar\sigma^{2}=2\sigma^{2}$. This is readily seen 
even in the crude approximation of $\bar W \sim \exp(-p^2/2\bar\sigma^2)$ 
and $W_0 \sim \exp(-p^2/2 \sigma_0^2)$, with 
$\sigma_0 \sim \exp(-\lambda t)$ and $t \gg 1/\lambda$. Therefore, when 
the purity starts decaying at the Lyapunov rate the echo does 
precisely the same. 

Using the above ideas we are now in a position to present a more complete
and illustrative picture of the time dependence of the echo $\bar M(t)$ 
and the purity $\varsigma (t)$. For the sake of simplicity we focus on the
echo but the same reasoning applies to the purity. To compute the 
overlap $\bar M=\int dxdp W_0\bar W$ we can split the phase space integral
into two regions: the region $A_C$ close to the classical unstable
manifold of the initial state, where $W_0$ is positive, and the region $A_O$ 
over which $W_0$ oscillates (see Figure 1):
\begin{equation} 
\bar{M}(t)=\int_{A_{O}} dx dp W_0 \bar W + \int_{A_{C}} dx dp 
W_0 \bar W. 
\label{Integral}
\end{equation} 
In the oscillatory region we can estimate the value of the integral 
assuming that there is a dominant wave vector $k_p$. In such a case, 
from Eq. (\ref{wignereq}) we can assume that 
$\bar W \simeq W_0 e^{-Dk_p^2 t}$. If more than one scale is present 
the result would be a sum of terms like this one. For the 
second integral, we can also use a 
crude estimate assuming that $W_0$ and 
$\bar W$ are constant over their respective effective support. In particular,
$W_0 \sim 1/A_C$ since its integral over $A_O$ cancels out. As $\bar W$ 
approaches the critical width $\bar\sigma$ along the stable manifold, 
the area of its effective support grows exponentially. Therefore, one 
gets that the second integral is 
$\int_{A_C}W_0\bar W\sim W_0 A_{C} \bar W \sim \bar W 
\sim e^{-\lambda t}$. Thus, combining the two results we find that the 
expected behavior of the Loschmidt echo is 
\begin{equation} 
\bar{M}(t) = a\ \exp(-\lambda t)+b\ \exp(-Dk_p^2 t)
\label{Msemiclassic} 
\end{equation} 
for appropriate prefactors $a$ and $b$. 

This result was previously derived for the Loschmidt echo using 
semiclassical techniques \cite{PastawskiJalab}. 
The first term gives the Lyapunov decay, while the second one 
describes the so--called Fermi golden rule regime (FGR) \cite{Jacquod}.
In this case the rate is proportional to the diffusion coefficient (which 
is itself proportional to the square of the strength of the perturbation).
As mentioned above, a similar result is expected for the purity. 

Our treatment  is valid in a semiclassical regime where the evolution 
of the Wigner function is dominated by the classical Hamiltonian flow 
and the corresponding interference fringes generated when its 
phase space support folds. The virtue of this 
analysis, entirely based on properties of the evolution 
of $\bar W$ derived in the context of decoherence studies, 
is not only its simplicity but also the fact that it enables us to 
identify the regions of phase space that can be associated with each of the 
terms appearing in (\ref{Msemiclassic}): the FGR contribution arises 
from the decay of the interference fringes while the Lyapunov contribution
is associated with the behavior of $\bar W$ near the classical unstable
manifold. Recent studies using semiclassical 
techniques point in the same direction \cite{CookPreparation}. 

It is interesting to perform a better estimate of the integral 
over the region $A_C$. Assuming that the local Lyapunov exponent is 
constant along the unstable manifold, one can approximate 
the value of the integral by using the corresponding result for the 
simplest system with an unstable fixed point: the inverted 
oscillator (IO) with Hamiltonian $H_0=p^2/2m-m \lambda^2 x^2/2$. 
In such a case, the echo can be computed exactly and turns out to be 
\begin{equation}
\bar M_{IO}(t)=\left(1+r\ {\rm sh}(2\lambda t)+r^2\left({\rm sh}^2(\lambda t)
-\lambda^2 t^2\right)\right)^{-1/2}.
\end{equation}
Here $r=\bar\sigma^2/4\sigma_i^2$, where $\sigma_i$ is the momentum 
dispersion of the initial state. This exact result shows that for 
long times ($\lambda t\gg \log(r)/2$) the echo $M_{IO}$ always decays 
as $\exp(-\lambda t)$. For initial times a decay 
with a rate determined by diffusion is observed but this transitory 
regime always leads to a decay dominated by the Lyapunov exponent. 
The initial transient is sensitive to the details of the noise 
statistics. For example, we can also exactly evaluate the echo when 
the noise kernel is flat (i.e., 
$\nu(t,t')$ independent of $t$ and $t'$). For such extreme case the 
long time behavior of the echo is not changed but the initial transient 
displays a quadratic decay.

We expect the analogy between Loschmidt echo and 
decoherence not only to enable intuitive derivations like the one
leading us to equation (\ref{Msemiclassic}) but also 
to provide new insights into theoretically unexplained 
experimental features such as the quadratic decay observed in 
\cite{PastawskiEnt}. 
Our results are also relevant for quantum computation as the Loschmidt 
echo is a measure of the fidelity with which a given 
algorithm is implemented. The Lyapunov decay of the fidelity could 
hinder the practical implementation of such computers, which would then 
have to deal with an exponential increase of error probability at a rate 
which is independent of the coupling to the environment and is 
solely fixed by the (possibly chaotic) nature of the underlying physical 
system. The high sensitivity of chaotic systems to perturbations seems to 
be connected to the efficiency of these systems to produce decoherence
on other systems, which has been under recent investigation 
\cite{ChEnviron}.

We thank Diana Monteoliva for help in producing the Figure. We also 
acknowledge support of ARDA/NSA grant. JPP received also partial support
from a grant by Fundaci\'on Antorchas.

\end{document}